RESEARCH ARTICLE

# Size agnostic change point detection framework for evolving networks

Hadar Miller☯, Osnat Mokryn☯*

Information Systems, University of Haifa, Haifa, Israel

☯ These authors contributed equally to this work.
* ossimo@gmail.com

**Data Availability Statement:** Enron dataset: Klimt B, Yang Y. Introducing the Enron Corpus. Machine Learning. 2004;. AskUbuntu dataset: http://snap.stanford.edu/data/sx-askubuntu.html.

**Funding:** OM received funding from the Israeli Science Foundation Grant 328/17. The funder had no role in study design, data collection and analysis, decision to publish, or preparation of the manuscript.

**Competing interests:** The authors have declared that no competing interests exist.

## Abstract

Changes in the structure of observed social and complex networks can indicate a significant underlying change in an organization, or reflect the response of the network to an external event. Automatic detection of change points in evolving networks is rudimentary to the research and the understanding of the effect of such events on networks. Here we present an easy-to-implement and fast framework for change point detection in evolving temporal networks. Our method is size agnostic, and does not require either prior knowledge about the network's size and structure, nor does it require obtaining historical information or nodal identities over time. We tested it over both synthetic data derived from dynamic models and two real datasets: Enron email exchange and AskUbuntu forum. Our framework succeeds with both precision and recall and outperforms previous solutions.

## Introduction

Complex systems of interacting elements, from human (social and organizational) to physical and biological ones, can be modeled as interaction networks, with nodes representing the elements and edges representing their interactions. When the interactions are dynamic, i.e., human and social interactions, a complete model that captures the longitudinal evolution of the system is comprised of a sequence of networks, each portraying a snapshot of the system at a single point in time [1, 2].

Of specific interest recently is the analysis of changes in dynamic social and complex networks in response to events, and the automatic detection of these points of change termed Change Point Detection (CPD). Recent works identified changes in the community partitioning of the Enron email exchange immediately after the Californian blackouts [3], and a turtling-up of conversation networks between traders in response to significant stock price changes [4]. Understanding the network's reaction to unusual events provides improved abilities to analyze, understand and possibly take actions in a given system, infer its response to external shocks, and aid in predicting organizational and behavioral changes.

Past research for identifying change points used stochastic models, of either scalar values representing the longitudinal data [5], or probabilistic and model-based representations of the network [3, 6–8]. However, the works mentioned above did not examine the complex network's structure as manifested through distributions.

The structural properties that are in the focus of our work here are the network's native statistical distribution, i.e., its degree distribution measure. Distribution functions are a measure of the division of resources within the network, and their relative positions, and are considered a fundamental tool in the understanding of complex systems. Stumpf and Porter [9] have discussed this notable role, claiming that degree distributions serve as an aiding tool for understanding, interpreting, and even predicting behaviors in a given system. Bhamidi *et al*. [10] further showed that degree-distribution measures reflect changes in the underlying structure better than the hyper-parameters of the corresponding network models. An additional valuable advantage of a degree distribution-based event detection is that it eliminates the need to know in advance the number of nodes in the network at each point in time and can work with as little information as the sequence of interactions for the periods under inspection. Thus, unlike other CPD schemes, the proposed solution here assumes no prior knowledge of the network, does not require pre-processing, and can be used in an online manner, where new network snapshots are generated on-the-fly.

Here, we devise an online fast change point detection mechanism, utilizing the degree distribution of snapshots of networks in time. The detection mechanism is planned in a manner that does not require to determine exact theoretical fits to the distributions. We conduct a hypothesis testing to assess the significance of the change and differentiate a change signal from local fluctuations. Our study here centers on interaction networks, such as phone calls, text messages, emails, and online social network postings. These networks, also termed temporal asynchronous human communication networks [11], can be characterized by the intertwining of the temporal topological structure and the interaction dynamics.

The contributions of the work are the following:

1. Taking a sliding-window approach for the network interactions, this method can address both the anomaly detection problem, in which there is a significant variation from a norm, and the change point detection problem, which considers a significant change to the norm itself, by computing the significance measure of the change (calculated p-value) over different window sizes.

2. The approach is the first that does not require to know in advance the number of interacting nodes in each stage of the network's life, and hence can be used online.

3. We investigate the performance of the scheme over both synthetic data and real-world data. For the synthetic data, we conduct a thorough investigation of several generative models, i.e., random networks and small-world networks, with varying rates of events and over different network sizes. This enables us to quantify the reaction of different network models to events. We further show that over two real datasets, the scheme performs better than existing detection schemes, while being faster.

4. The hypothesis testing we conduct enables a sensitivity measure for a change. First, it enables to set the level of sensitivity of a change according to need. Then, it opens the possibility to detect changes with decreasing sensitivity during a window of time. While current schemes detect reactions to shocks, this scheme can detect gradual changes that follow a clear trend of increasing probability of a change and can be utilized as a predictive framework.

## Related work

In this work, we investigate the effect of events on social networks. Romero, Uzzy, and Kleinberg in a recent novel work [4] defined these events as mostly exogenous events that are either unexpected or are extreme, relative to the average [4, 12]. They found a turtling-up of the network as a reaction to an external shock, and measured changes in the clustering coefficient, tie strength and percentage of border edges. Kondor *et al.* [13] researched the longitudinal structure of the network of the most active Bitcoin users for two years and searched for important changes in the graph structure by comparing successive snapshots of the active core of the transaction network using principal component analysis (PCA). They found a clear correspondence with the market price of Bitcoin. McCulloh & Carley [5] included in their study of change points also cases of endogenous changes and showed that their detection system could determine that a change has occurred from a longitudinal analysis of the network itself. Using their method, Tambayong [14] examined Sudan's political networks and found that foreign-brokered signings of multiple peace agreements served as a political solidification point for political actors of Sudan during the recent violent domestic conflict. According to their analysis, this was a catalyst that caused three leaders to have emerged and lead the more compartmentalized yet faction-cohesive political networks of Sudan. Peel and Clauset [3] were able to detect external changes during the Enron crisis through a stochastic analysis of the Enron organizational email exchange [15].

In stochastic models of networks, change points are points in time where a change in the system's norm is detected in a manner that can be significantly differentiated from plain stochastic noise [3, 16–19]. McCulloh & Carley [5] convert the series of networks to a time series of scalar values for different network measures, looking for a stable change in these values (as opposed to temporal change, when looking for anomaly detection) using process mining techniques for change points detection [5, 20, 21].

Methods for CPD differ mainly by the graph features they compute. A model-based approach fits each snapshot to a generative model. Examples for models used are the General Hierarchical Random graph (GHRG), Generalized Two Block Erdos-Renyi (GBTER), and Kronecker Product Graph Model (KPGM) [3, 22, 23]. Structural entropy is a new measure suggested by Almog and Shmueli [8] that takes into account the number of communities and their sizes, and is a revised interpretation of the Structural Diversity metric. It was utilized to monitor changes in the structure of correlation-based networks over time. A model-based approach requires a pre-processing phase, for which enough history is pre-known. Some works further require that labeled nodal information is known. When taking the degree distribution, we eliminate the need for this extended information, as degree distribution does not require historical information, nor the node names. Moreover, a recent analysis found that the degree distribution better detects structural changes than the hyper-parameters of the generative model for the PA case [10].

A complementary approach, similar to ours, is to extract a large number of features from each consecutive graph snapshots and find the distance between them [6, 7, 16, 24]. A change is determined if a predefined threshold for the distance is crossed. Anomaly detection analogous techniques suggest eigenvalue-based detection [25] and matrix factorization [26], to name a few. The use of the Relative Hausdorff (RH) distance metric for anomaly detection in temporal networks with long-tail distance distribution was recently discussed in [27]. Indeed, different distance metrics can be used in our framework. We later discuss in more detail the differences between distance metrics, including RH, and suggest when to use which.

In this work, we focus on temporal asynchronous human communication networks [11]. Previous research by Braha and Bar-Yam found, surprisingly, that "the existence of a link between two individuals at one time does not make it more likely that the link will appear at another time" [28, 29]. They found a weak correlation between consecutive networks of interactions, while the degree distributions of these networks remained similar. Further, Palla *et al.* [30] found that while small groups can stay stable in time if their membership is constant, the opposite is true for larger communities: "The condition for stability for large communities is continuous changes in their membership, allowing for the possibility that after some time practically all members are exchanged." These findings motivate our choice of utilizing the networks' degree distributions for change detection.

Unlike previous works that consider graph features, in our work, we conduct a hypothesis test to provide certainty for a change point detection. When used gradually, it can be further used to detect an incremental change.

## Method

We explain our method, following Fig 1. A sequence of networks is presented, where a change in the generative model occurs. The change is not tied to a specific structural characteristic. Our framework computes the cumulative distribution function of the degrees (CDF) for each graph, computes the distance, and performs a hypothesis testing to infer how probable is a change given the measured distance between the two CDF's. Here, we chose the non-parametric Kolmogorov-Smirnov (KS) two-sample test to measure the distance, as discussed later. Other non-parametric statistical methods for measuring the distance may be applied. We discuss alternatives in the Results Section.

Frameworks for change point detection divide the data to consecutive snapshots according to a natural division derived from the nature of the data, such as daily or weekly snapshots of organizational frameworks, or monthly graphs of votes. In methods measuring the distance between features extracted from two consecutive graph snapshots [6, 7, 16], a change is

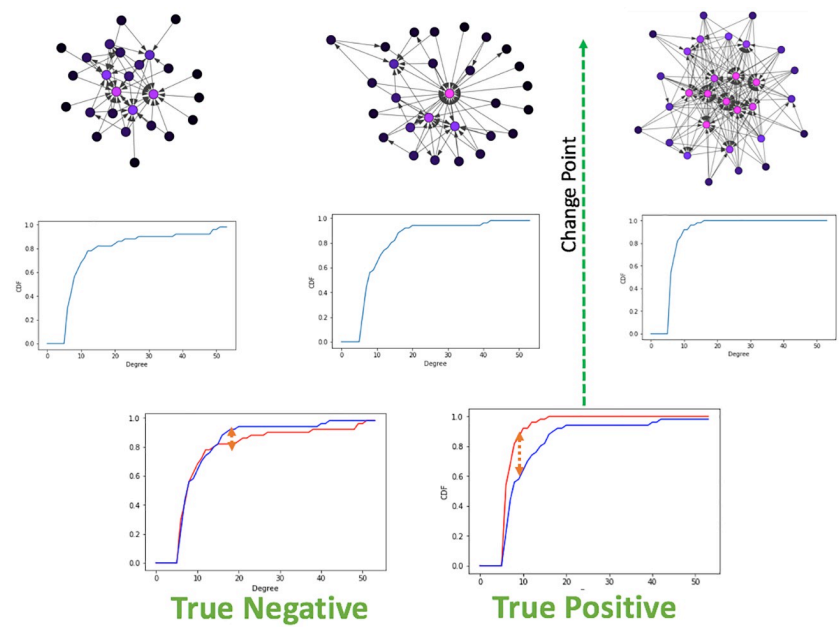

**Fig 1. Framework for detection of changes.** Our Window size defines the stability of the change over time. Hypothesis testing over a distance measure is used to determine whether the underlying model has changed. On the left graphs generated from the same model, on the right a graph generated from a changed model.

https://doi.org/10.1371/journal.pone.0231035.g001

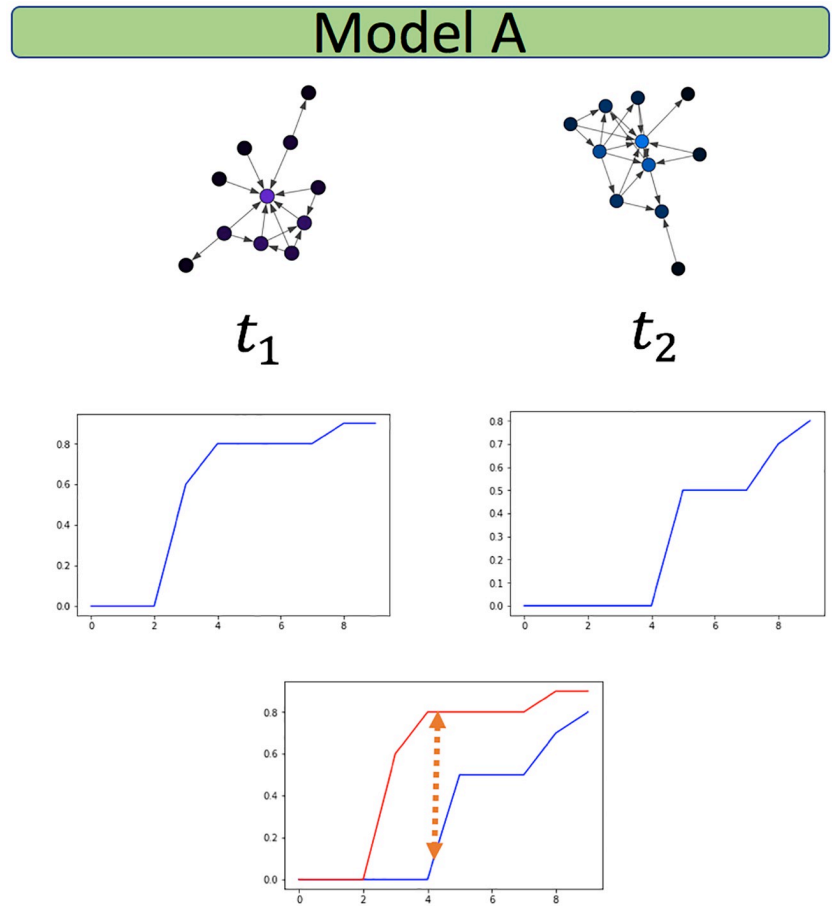

**Fig 2. False positive.** Distance measure is large as the sample size is too small, although graphs come from the same generative model.

https://doi.org/10.1371/journal.pone.0231035.g002

detected if the measured distance is bigger than an arbitrarily predefined threshold value. A drawback of distance measures is that they work well mainly for large sample sizes. When the sample size is small, a large distance can be measured, crossing the predefined threshold value. This large measure can lead to a false-positive result, i.e., an inference of a change when there is merely a fluctuation. Fig 2 demonstrate such a false-positive scenario. To avoid these types of false inferences, we suggest the use of a sliding window over several graph snapshots and computing the CDF across the entire window, as is the case in Fig 4. A complimentary situation occurs when windows that are set too large, conceal an event within them, thus hiding the point of change. This would correspond with a false negative inference and is demonstrated in Fig 3. A solution to this problem is the use of a sliding window to find the exact point of change within the window, as is used in [3].

An alternative approach to measuring a distance between windows is to try and fit a theoretical statistical distribution to each network snapshot, and determine whether they are derived from the same model. This is, however, a rather time and computational-intensive approach. Fitting data to a statistical theoretical model requires both to find a fit and to reject other possible theoretical statistical distributions [31]. Hence, we compare distances across windows, as described in Fig 4, and perform a hypothesis testing of a change.

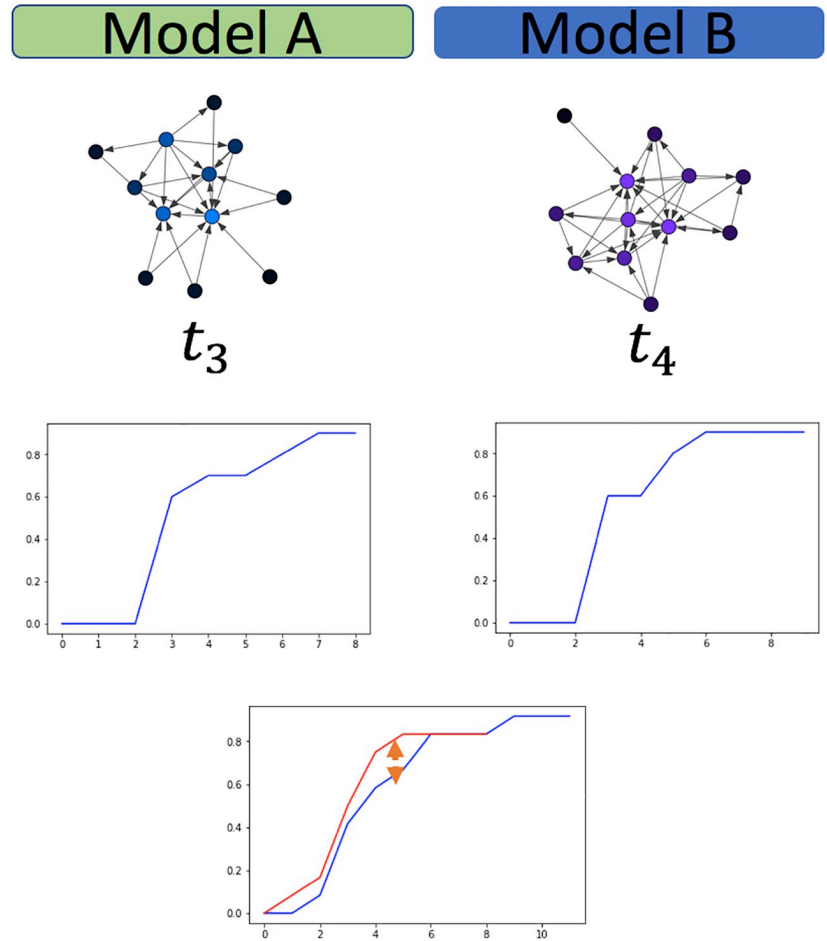

**Fig 3. False negative.** Fluctuations conceal each other and decrease the measured distance between two networks.

https://doi.org/10.1371/journal.pone.0231035.g003

We conduct a hypothesis testing to learn whether the distance between the degree distributions affirms that they come from the same model or two different generative models. We measure the distance between the cumulative degree distributions of consecutive snapshots. For any two consecutive windows, let us define their graphs representations as $g_i$, $g_{i+1}$. The null hypothesis is that the cumulative distributions measured for any two consecutive snapshots, $g_i$, $g_{i+1}$, are drawn from the same distribution, $G_{Null}$, in which case no change has occurred between the windows. To test the hypothesis, we generate synthetic datasets from the distribution of $g_i$ and find their distributions. The standard approach for generating samples for hypothesis testing is a Monte-Carlo bootstrap process, which generates samples by randomly re-sampling (with replacement) the data [32].

We proceed to explain the process in more detail. We use here the non-parametric Kolmogorov-Smirnov (KS) two-sample test. The method is considered robust and is widely used (as discussed at the end of this section further). Yet, when comparing two distributions using too few samples, it can fail to reject a false null hypothesis. The probability for that diminishes as the number of nodes interacting in each snapshot increases.

For two consecutive graph snapshots $g_i$, $g_{i+1}$, $(i \in \{1, 2, \ldots\})$ we denote the two generated corresponding cumulative degree distribution functions by $S_i(x)$, $S_{i+1}(x)$. Given the CDF degree distribution $S_j(x)$, $j \in i, i+1$ for graph $g_j$: $S_j(x) = P_j(x \leq X)$ we compute the KS statistic

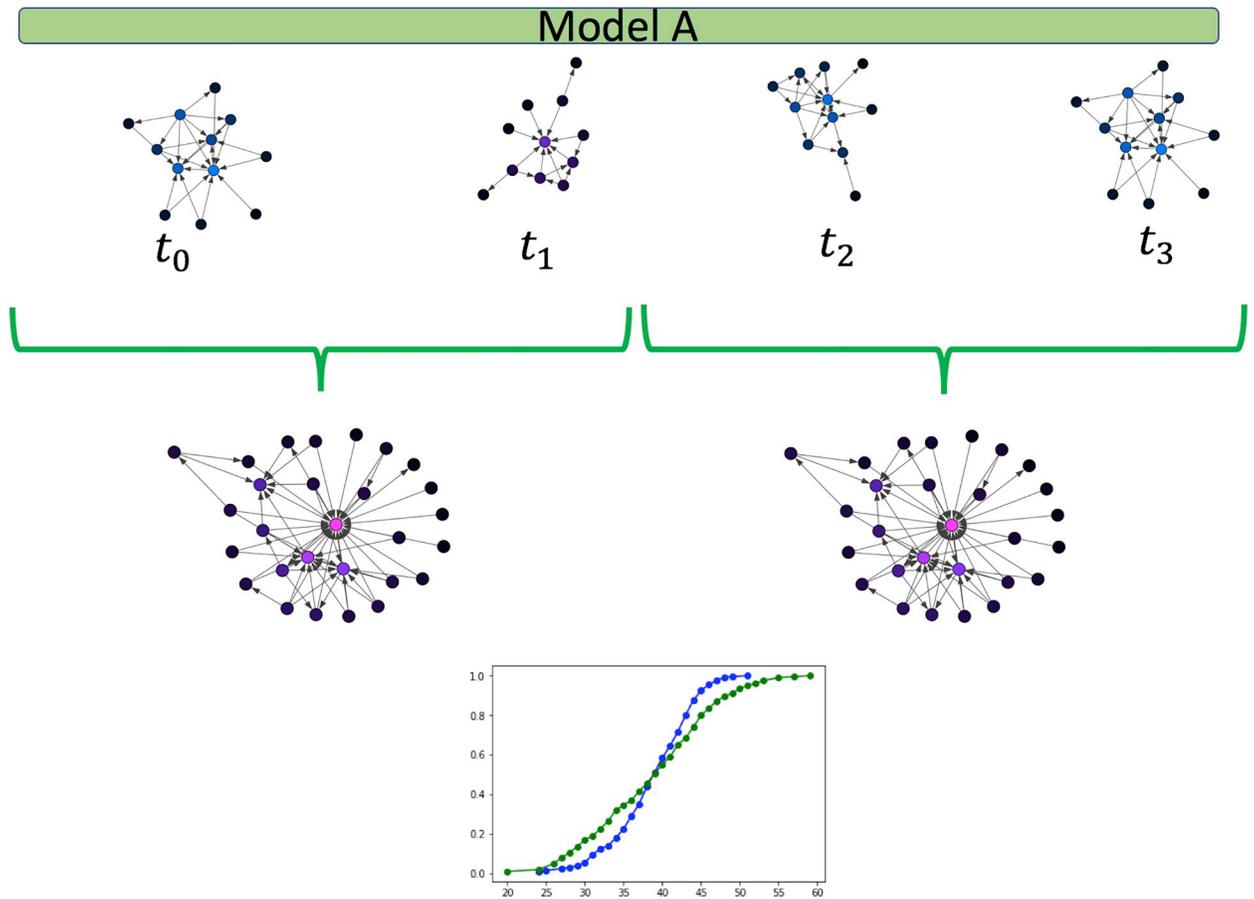

**Fig 4. Sliding window.** The use of a sliding window over several graph snapshots decreases the probability of a false positive estimation of a change.

https://doi.org/10.1371/journal.pone.0231035.g004

$D$, defined as the maximal difference between the two empirical distributions, as described by Eq 1. The KS null hypothesis is that the two samples were drawn from the same distribution.

$$D(S_i, S_{i+1}) = \sup_x |S_i(x) - S_{i+1}(x)| \qquad (1)$$

The KS null hypothesis is rejected if the computed distance $D(S_i, S_{i+1})$ is greater than some critical value, usually a predefined threshold.

In our framework, we suggest a hypothesis testing that determines for each distance $D(S_i, S_{i+1})$ the confidence with which the null hypothesis can be rejected for it, as follows. As explained before, a large KS distance $D(S_i, S_{i+1})$ measured between $S_i(x)$ and $S_{i+1}(x)$ doesn't necessarily indicate that the model has changed. We would like to test how rare such distance ($D(S_i, S_{i+1})$) is, given $S_i(x)$. To do so, we define $g_i$ as the base model graph and conduct a hypothesis testing, with a null hypothesis that the distance $D(S_i, S_{i+1})$ between the base model graph distribution and the consecutive one, $g_{i+1}$'s distribution, is not rare for samples taken from the same statistical model. Our null hypothesis then assumes that the distance between the model graph distribution CDF, $S_i(x)$, and the consecutive graph's distribution CDF, $S_{i+1}(x)$, is typical for distances between distributions sampled from the same base model graph distribution. The null hypothesis is rejected with significance $p$ if in $\alpha = (1 - p)$ of the times the

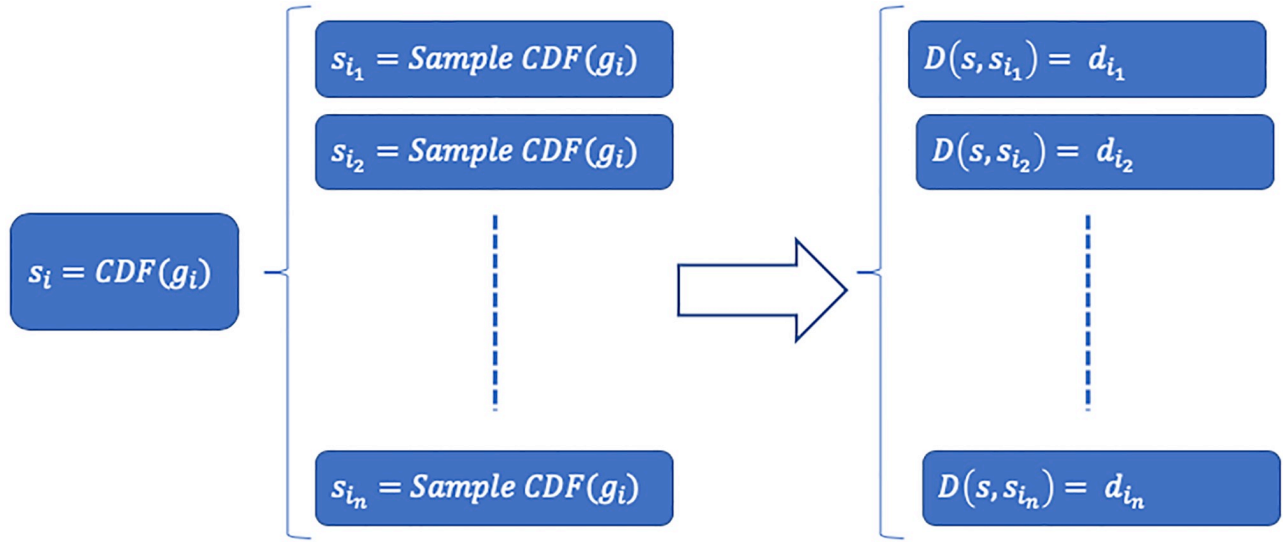

**Fig 5. Monte-Carlo bootstrap procedure:** $S_i$ is sampled $n$ = 1000 times (with repetitions) to create the series $S_{i_k}, k \in [1..n]$. Then, we measure the KS distance from $S_i$ to each of the sampled CDFs, yielding the group of distances $\{d_{i_k}\}, k \in [1..n]$.

https://doi.org/10.1371/journal.pone.0231035.g005

measured distance between $S_i(x)$ and the sampled distributions is smaller than $D(S_i, S_{i+1})$, as depicted in Eq 2.

Following the Monte-Carlo bootstrap procedure [32] we generate $k$ = 1000 >>1 new samples from $S_i$ and compute the distance $D(S_i, S_{i_k})$ between $S_i(x)$ and each of its bootstrap samples, $\{S_{i_k}(x), k \in 1..1000\}$. This results in a group of size $n$ = 1000 of distances from $S_i$ to its samples, namely the group of distances $\{d_{i_k}\}, k \in [1..n]$. Fig 5 gives a visual description of this process. Then, given $k \in [1..n]$:

$$\forall \alpha < 1, \ \exists \ d_{i,\alpha}, \ \text{s.t.} \ |\{D(S_i, S_{i_k}) < d_{i,\alpha}\}| = \alpha \cdot n \qquad (2)$$

Namely, there exists a distance $d_{i,\alpha}$ that is greater than the distance between $S_i(x)$ and $\alpha$ of its bootstrap-generated samples. If $D(S_i, S_{i+1}) > d_{i,\alpha}$ then we can reject the null hypothesis with confidence $\alpha$.

For example, let us assume that $\alpha$ = 0.95. Then, following the definition in Eq 2, $d_{i,0.95}$ is greater than the distances between $S_i(x)$ and 95% of its bootstrap-generated samples. If, then, the distance between $S_i(x)$ and the subsequent graph's distribution, $S_{i+1}(x)$, $D(S_i, S_{i+1})$, is greater than $d_{i,0.95}$, then we can reject the null hypothesis and claim with confidence of 95% that $g_i, g_{i+1}$ are graphs that are generated by difference models.

Next, we continue to evaluate the performance of our framework over both synthetic and real-world networks and discuss alternatives to the KS distance metric.

## Results

We conduct several experiments to demonstrate the performance of our framework on both large synthetic datasets and real networks. First, we investigate the performance of the framework on sampled synthetic networks generated by known generative models and then follow a rigorous evaluation to evaluate the sensitivity of our framework with the suggested distance metric over synthetic datasets. We then continue to evaluate the framework over two real datasets and compare our results against a known baseline solution. We conclude this section by evaluating the performance of different distance metrics that can be used with our framework and discuss their differences.

### Synthetic datasets

We conduct a series of experiments over synthetic datasets generated by known generative models. Each such generative model enables us to investigate the framework's behavior for different structural characteristics. As our method is based on the degree distribution of the network, it is agnostic to any changes in the network size. Hence, we expect our framework to detect changes across network snapshots that may gain or lose nodes during the network's lifetime. At first, we considered using a preferential-attachment growing network as one of the models. However, this model is specifically designed to explain the emergence of hubs in networks and the long tail distribution of real-world network degrees, and thus is designed to create a specific degree distribution, which is what we try to find, resulting in a trivial test. For generative models, we, therefore, employ the Erdös-Rényi (ER) random networks model and the Caveman model. First, we conduct three experiments over these networks, and then we perform a large scale sensitivity test for both models.

**Designed experiments.** In each experiment, the network model alternates between two configurations that differ in their hyper-parameters. The number of changes is set to 100, distributed randomly. Then, the number of consecutive snapshots of the network drawn from the model configuration, $x$, is chosen from a normal distribution $x \sim N(\mu = 4, \sigma^2 = 2)$, such that each point is the averaged results of 400 runs.

**Random graphs.** We start with the Erdös-Rényi (ER) random graph model [33]. The model for random graphs $G(n, p)$ assumes a fixed number of nodes $n$ (Recently Zhang et al. [2] suggested a generalization for dynamic random networks, in which the dynamic process is governed by a continuous Markov-process. As we need to study the change point detection problem, requiring a change in the generative model hyper-parameters, we could not employ their model.) Edges connect node pairs independently with probability $p$. Low values of $p$ entail that the number of edges is substantially lower than the number of nodes and the model generates small components in tree forms. As $p$ increases, and reaches $p > o\left(\frac{1}{n}\right)$, the network changes to suddenly form a giant component, a phase transition that has a distinct influence on the structure of the network.

We then perform two experiments for this model type, as described here, and detailed in Figs 6 and 7:

- **Experiment 1**: A change in the hyper-parameters of the ER model transitions the network between the two network states of fragmented ($p << o\left(\frac{1}{n}\right)$) and connected ($p > o\left(\frac{1}{n}\right)$). The network configurations are the following. Each configuration consists of 200 nodes, and the model's hyper-parameter is either $p = 0.003$, i.e., fragmented, or $p = 0.01$, i.e., connected.

- **Experiment 2**: The ER networks consist of 200 nodes each, and the model's hyper-parameter is either $p = 0.1$, $p = 0.15$, i.e., both times the network is connected, and there is a slight change in the connectedness. It is safe to assume that the subtleness of the change in the generative model of the random network will make it harder to identify.

**Caveman model.** The ER model generates a graph with small clustering coefficients, which lack the capability to represent communities. Social networks are often characterized as having highly connected communities that form rare interactions in between and form a *Small World*. For example, in an organization, one may expect intensive interactions between actors within departments and sparse interactions between actors belonging to different departments. A generative model for a small-world network is the Caveman [34].

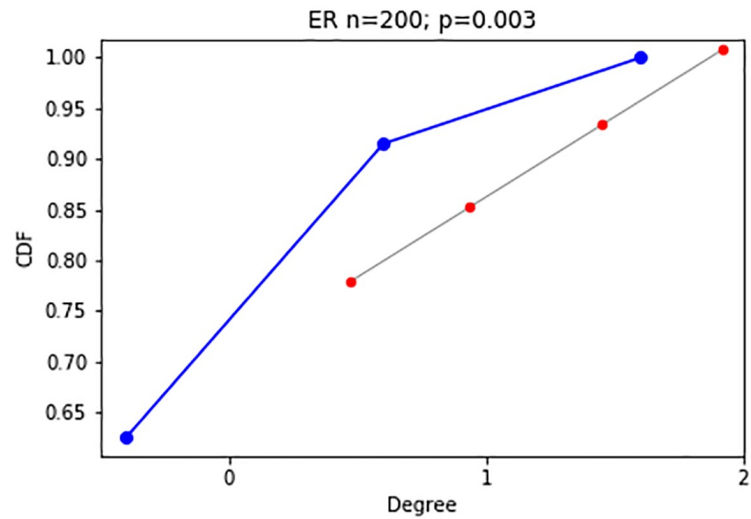

**Fig 6. Random networks: Experiment 1 (200 nodes).** Experiment 1 1st configuration ($p = .003$, fragmented) degree CDF, visualization of largest component.

https://doi.org/10.1371/journal.pone.0231035.g006

To test our framework against networks with varying sizes we generated a sequence of unlabeled networks, $g_i \in G$, while using the Caveman model. The number of nodes for each snapshot was randomly selected from a uniform distribution $\|g_i\| \sim U(200, 1000)$. To prevent a sample size bias while calculating the KS distance we randomly sampled 200 nodes from each network and calculated the distance between the two samples degree distribution.

- **Experiment 3**: The Caveman-based networks are drawn from a model containing 200 nodes, as explained above, and $C = 5$ communities each. The change in the hyper-parameter

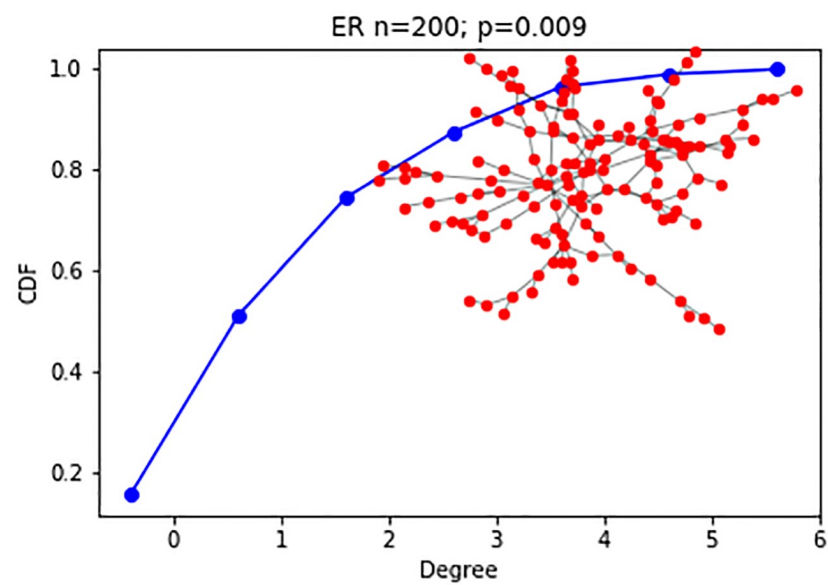

**Fig 7. Random networks: Experiment 1 (200 nodes).** Experiment 1 2nd configuration ($p = .009$, connected) degree CDF, visualization of largest component.

https://doi.org/10.1371/journal.pone.0231035.g007

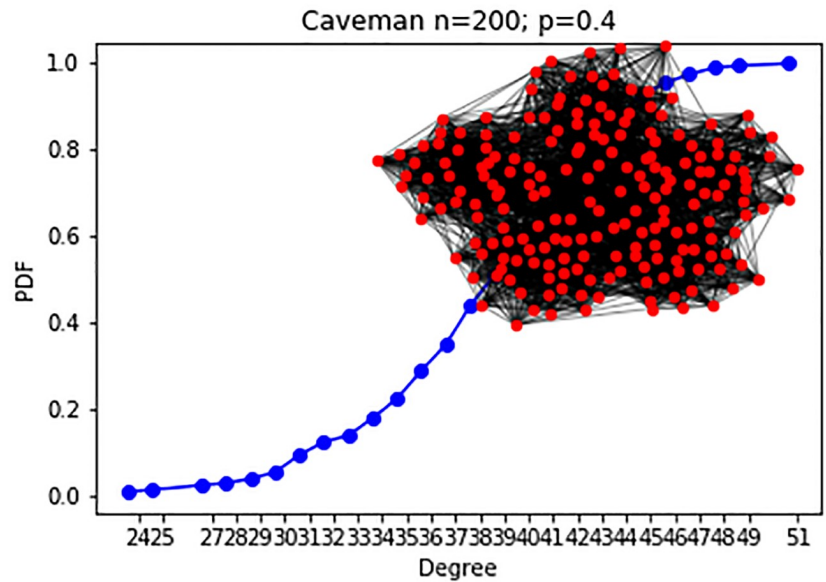

**Fig 8. Caveman: Experiment 3 (200 nodes, C = 5).** Experiment 3 1[st] configuration ($p = 0.4$) CDF, visualization.

https://doi.org/10.1371/journal.pone.0231035.g008

between the two configurations is in the rewire probability $p$. In the 1[st] configuration, visualized in Fig 8 $p = 0.4$. In the 2[nd] $p = 0.7$, leading to a more inter-connected network, as is visualized in Fig 9.

**CPD performance for the synthetic networks examples.** Table 1 describes the performance of our detection framework for the three described experiments. Note that for the ER networks (exp1, exp2), we get a perfect recall. The degree distribution of an ER random graph

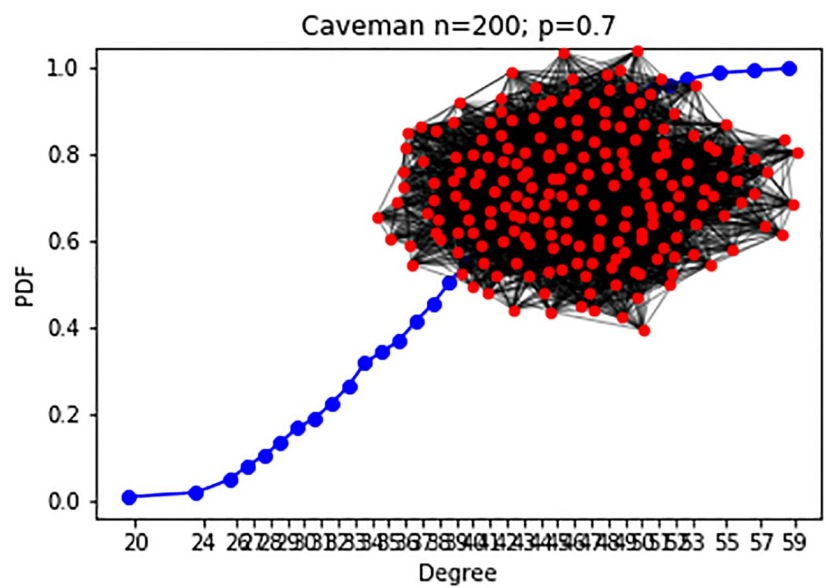

**Fig 9. Caveman: Experiment 3 (200 nodes, C = 5).** Experiment 3 2[nd] configuration ($p = 0.7$) CDF, visualization.

https://doi.org/10.1371/journal.pone.0231035.g009

**Table 1. CPD framework performance for synthetic networks examples.**

| Experiment | Model & Main Structural Property | Precision | Recall |
|---|---|---|---|
| | | Mean, Std | Mean, Std |
| exp1 | ER: Phase transition $p$ = {0.003, 0.009} | 0.767, 0.03 | 1.0, 0.0 |
| exp2 | ER: Connected $p$ = {0.1, 0.15} | 0.671, 0.02 | 1.0, 0.0 |
| exp3 | Caveman: Communities $p$ = {0.4, 0.7} | 1.0, 0 | 0.961, 0.01 |

https://doi.org/10.1371/journal.pone.0231035.t001

with edge probability $p = \frac{\lambda}{n}$ follows a Poisson distribution with probability mass function: $e^{-\lambda}\frac{\lambda^k}{k!}$, with mean $\lambda$ and skewness $\lambda^{-0.5}$. A change in $\lambda$ differentiates two ER generative models and will be projected to the networks' CDF, thus detectable by our model. This may explain the perfect detection (Recall = 1) of all events in our synthetic data tests. However, the variance of a Poison distribution is $\lambda$ as well. As the variance $\lambda$ increases, the chances of mistakenly find two samples drawn from the same model as not sharing the same distribution increase. This explains our relatively low precision.

As true positive events (change points) were detected with significance that exceeds 99% We repeated the experiments while increasing the CPD threshold from 90% to 99%. This test resulted with Recall = 1.0 and Precision = 0.89. This corresponds to changing the *sensitivity* of the framework, as discussed before.

The results for the Caveman model (exp 3) yield excellent results of perfect precision (100%) and near-perfect recall (96%), showing that a community structure of networks lends itself naturally to our detection framework.

**CPD framework's performance—Sensitivity test.** We continue to understand the sensitivity of the framework, given the KS metric (As noted before, we will examine other distance metrics' performance in a follow-up section.) to changes in the models' hyperparameters. We examine both models as before, the ER random networks, and Caveman. The ER network's hyperparameter, $p_{ER}$, is the edge probability, and thus dictates the network's density. The larger $p_{ER}$, the larger the network's average degree. As for the Caveman model, the corresponding hyperparameter, $p_{CV}$, indicates the amount of rewiring and is thus inversely correlated with the network's cluster-coefficient. The temporal networks' sizes are randomly picked from a normal distribution: $n \sim N(\mu = 100, \sigma^2 = 10)$. The Normal distribution with these parameters was chosen as it represents the sizes of the temporal weekly networks of real-world datasets used in this research.

Fig 10 depicts the results of the $f_1$ measure of the performance of our framework with the KS distance metric for all permutations of ER networks that entail change. The networks were modeled with the edge connectivity probability, $p_{1,ER}, p_{2,ER} \in [0.05, 0.1, 0.15, \ldots 1]$, $p_{1,ER} \neq p_{2,ER}$. Overall, the sensitivity graph depicts 380 experiments, in which each point is the average of 2x100 random networks. The framework excels in finding the hyper-parameter change. Hence, it is very sensitive to changes in the network's density, and fails to find a change only when the changes are very small, i.e., $|p_{1,ER} - p_{2,ER}| = 0.05$, as can be seen by the low $f_1$ value at the narrow diagonal line.

We repeated the experiment above for the Caveman model, depicted in Fig 11. The networks were generated with the rewiring parameters $p_{1,CV}, p_{2,CV} \in [0.05, 0.1, 0.15, \ldots 1]$, $p_{1,CV} \neq p_{2,CV}$. Similarly, each point is the result of 100 experiments. The rewiring parameter in the model correlates negatively with the network's clustering coefficient, as it determines the probability of edge rewiring. As the network is formed with a low number of communities that are fully intra-connected, higher rewiring probability corresponds to lower values of the clustering

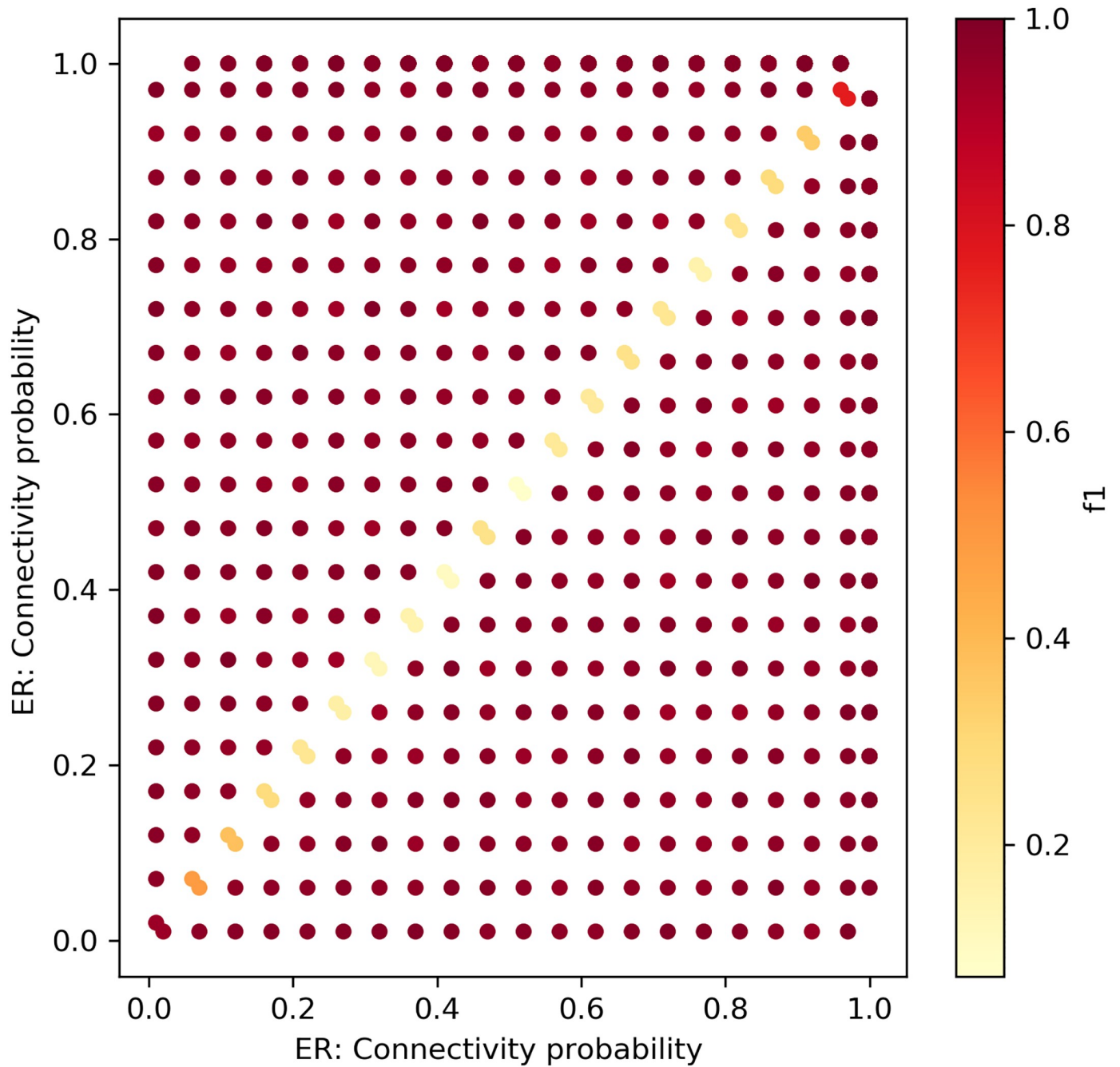

**Fig 10. The framework with KS distance—ER networks.** Sensitivity test for ER networks with sizes $\sim N(\mu = 100, \sigma^2 = 10)$.

https://doi.org/10.1371/journal.pone.0231035.g010

coefficient. The results of our framework with the KS distance metric are moderate. It is not sensitive to small changes (fails to find a change when $|p_{1,CV} - p_{2,CV}| = 0.1|$), and as the $p_{CV}$ value increases, the framework's ability to detect changes decreases. Hence, we can conclude that the framework with the KS metric fails to detect changes between networks that exhibit low clustering coefficient values.

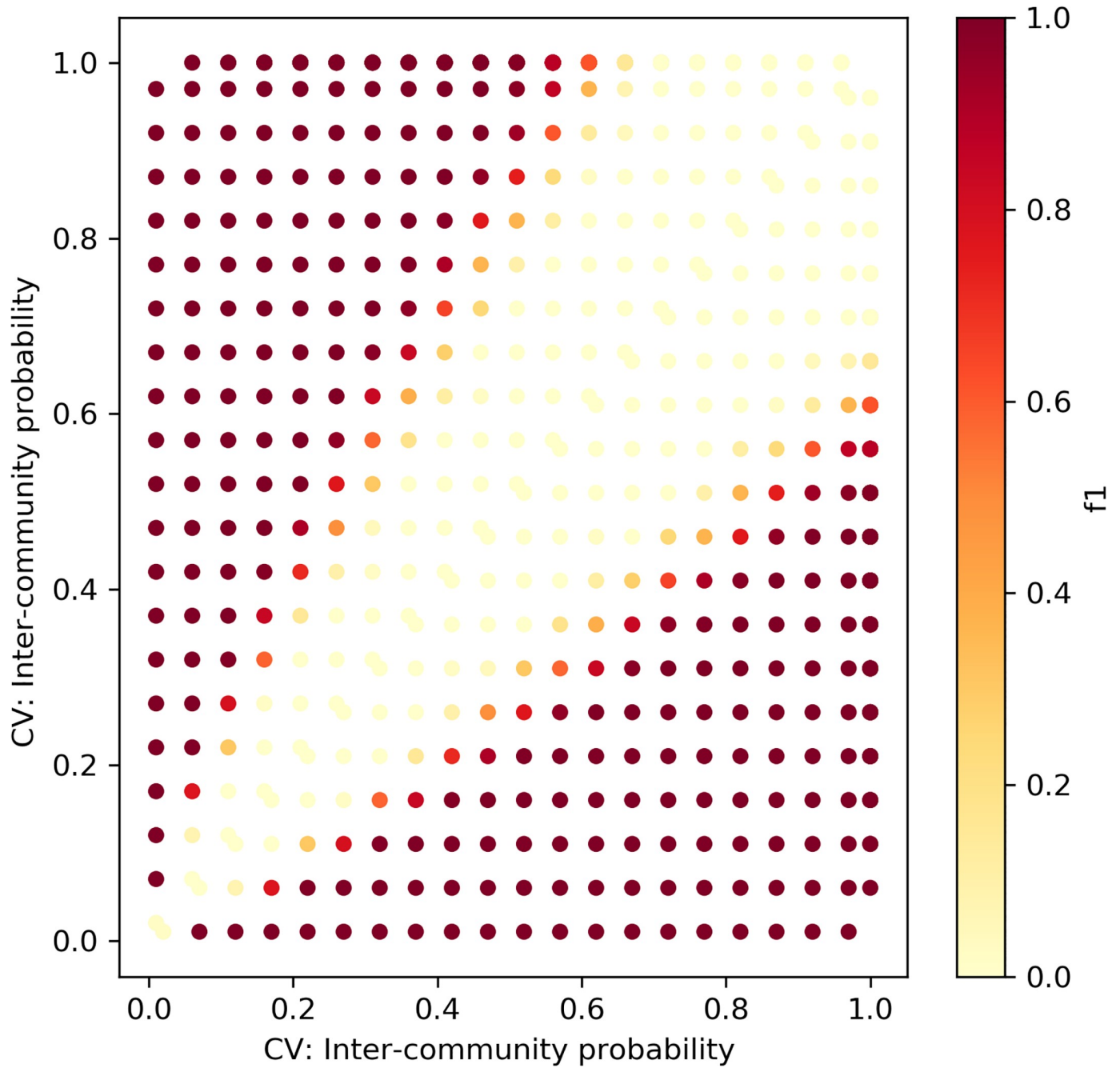

**Fig 11. The framework with KS distance—Caveman networks.** Sensitivity test for Caveman networks with sizes $\sim N(\mu = 100, \sigma^2 = 10)$.

https://doi.org/10.1371/journal.pone.0231035.g011

### Real-world datasets

We tested our framework against two real-world datasets. The first, the Enron email exchange between 151 employees, mostly managers [15]. We generated weekly networks from the email interactions similar to [3, 7].

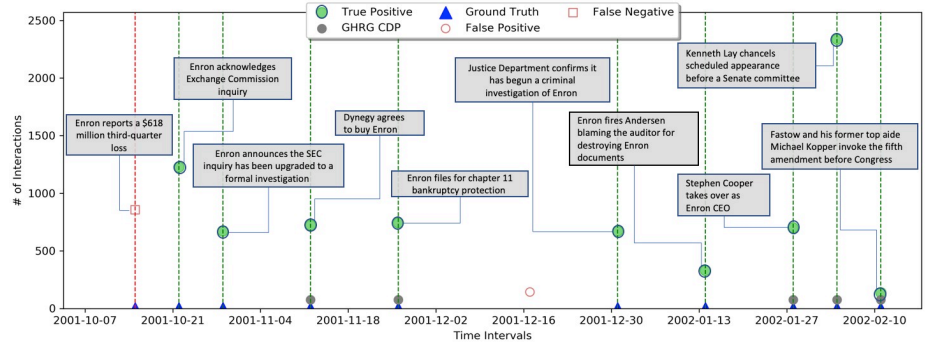

**Fig 12. Enron emails exchange.** The Enron email exchange during the second half of 2001, where many events took place. Real events denoted by blue rhombuses, True positive detection events by a green star. In grey, the events detected by the GHRG framework, which is used as a baseline for comparison. Our framework outperforms with Recall = 0.9 and Precision = 0.9.

https://doi.org/10.1371/journal.pone.0231035.g012

Fig 12 describes our framework's performance, compared to both the real events, and to the GHRG-based detection framework by Peel and Clauset [3]. Our framework detected 13 out of 14 change points, resulting in recall and precision of 0.9.

The second dataset is the interactions on the stack-exchange website AskUbuntu [35]. The periods are of months. We assume that a new Ubuntu release might affect the community, and extracted the release dates from the Ubuntu site. The number of interacting nodes (participants) varies greatly in the measured period, from 4095 to 6202, as is the amount of interactions. As described before, our method allows for different participants and interactions between windows while measuring the distribution. The results are described in Fig 13. Our framework detects almost all the events with high confidence: Recall = 0.8, and Precision = 0.57. There might be external events (i.e., are not version releases) that we are not aware of that could have affected the network as well.

## Considerations in distance metric choices

The KS test statistic uses the difference of greatest magnitude between two distributions and is more sensitive to discrepancies between the distributions around the middle. Additionally, it

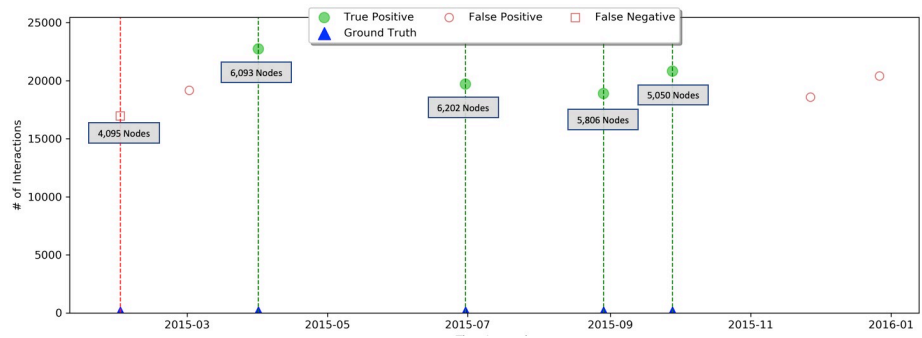

**Fig 13. AskUbuntu forum exchange.** Release events denoted by blue rhombuses; True positive detection events by a green star. Our framework performance yields Recall = 0.8 and Precision = 0.57, number of nodes in each window is denoted on the graph per detected event.

https://doi.org/10.1371/journal.pone.0231035.g013

is efficient, fast to compute, and was found useful for change detection in data streams [36]. In the case of moderate-sized asynchronous human communication networks of the type examined in our work, the KS distance metric yields very good results. However, Granger [37] further showed that a large amount of data might not be advantageous for inferential statistics such as the KS test. When the temporal degree in the different snapshots follows a long-tail distribution, a distance metric that takes into account small changes in the tail of the distribution would be preferred. Examples for such networks are the Kullback-Leibler divergence (KL) [38] and Relaxed Hausdorff (RH) [27] distance metrics. The RH is a relative new distance metric, and was found to outperform the KL distance metric. It is defined as follows. Two graphs $G$, $F$ are defined to be $(\epsilon, \delta)$ – close if:

$$\forall d, \exists d' \in [(1-\epsilon)d, (1+\epsilon)d], \text{s.t.} |F(d) - G(d')| \leq \delta F(d) \tag{3}$$

Where if $F$, $G$ are $\epsilon$ – close then RH(F,G) = $\infty$. The metric is defined in a manner that does not limit the distance, hence the threshold is data-driven. It is tailored to find small discrepancies between long-tailed streams for anomaly detection and requires that enough samples would exist in each graph for change detection. In our real datasets, this is not the case. For example, in the Enron dataset, there are around a hundred nodes in each of the snapshots. As RH does not work on the CDF of the graph, it requires padding if the number of nodes differs between the measured graphs. Still, the RH metric outperformed all other metrics for change point detection in long-tail networks that we checked. We continue here to discuss the performance of RH and KL for moderate-sized networks.

Figs 14 and 15 depict the change detection performance of the RH and KL distance metrics, correspondingly, in detecting changes between ER networks of sizes $\sim N(\mu = 100, \sigma^2 = 10)$. Recall that the normal distribution with these parameters was chosen as it represents the majority of the weekly temporal network snapshots of the real-world datasets we use in this research. When the changes are small, both metrics fail to detect changes. However, when the density increases, KL can better detect small changes between ER networks. Figs 16 and 17 depict the change detection performance of the RH and KL distance metrics, correspondingly, in detecting changes between Caveman networks of sizes $\sim N(\mu = 100, \sigma^2 = 10)$. KL performs well and can detect rather small changes. RH is tailored for detecting changes between networks with a long-tail degree distribution and thus does not perform well for networks of small size.

## Results summary

Table 2 Summarizes the framework's performance compared to the relevant competition for the real-wold dataset for which we had ground truth, and for the synthetic networks sensitivity tests. For the real dataset (Enron) our framework found 13 out of the existing 14 points of change in the data, and achieved recall and precision of 0.9, compared to the GHRG-based solution suggested in [3] that did not have false positives but failed to identify some of the change-points. We further compared our framework performance over the synthetic datasets to the performance of two alternative distance metrics between the degree distributions of networks: KL and RH. For both random networks (ER) and Caveman-based networks of mid-sizes, our framework outperformed the competition.

## Discussion

It is widely accepted that the structural properties of a network play a significant role in determining its actors' behavior [39–44]. The last decade's abundance of temporal information

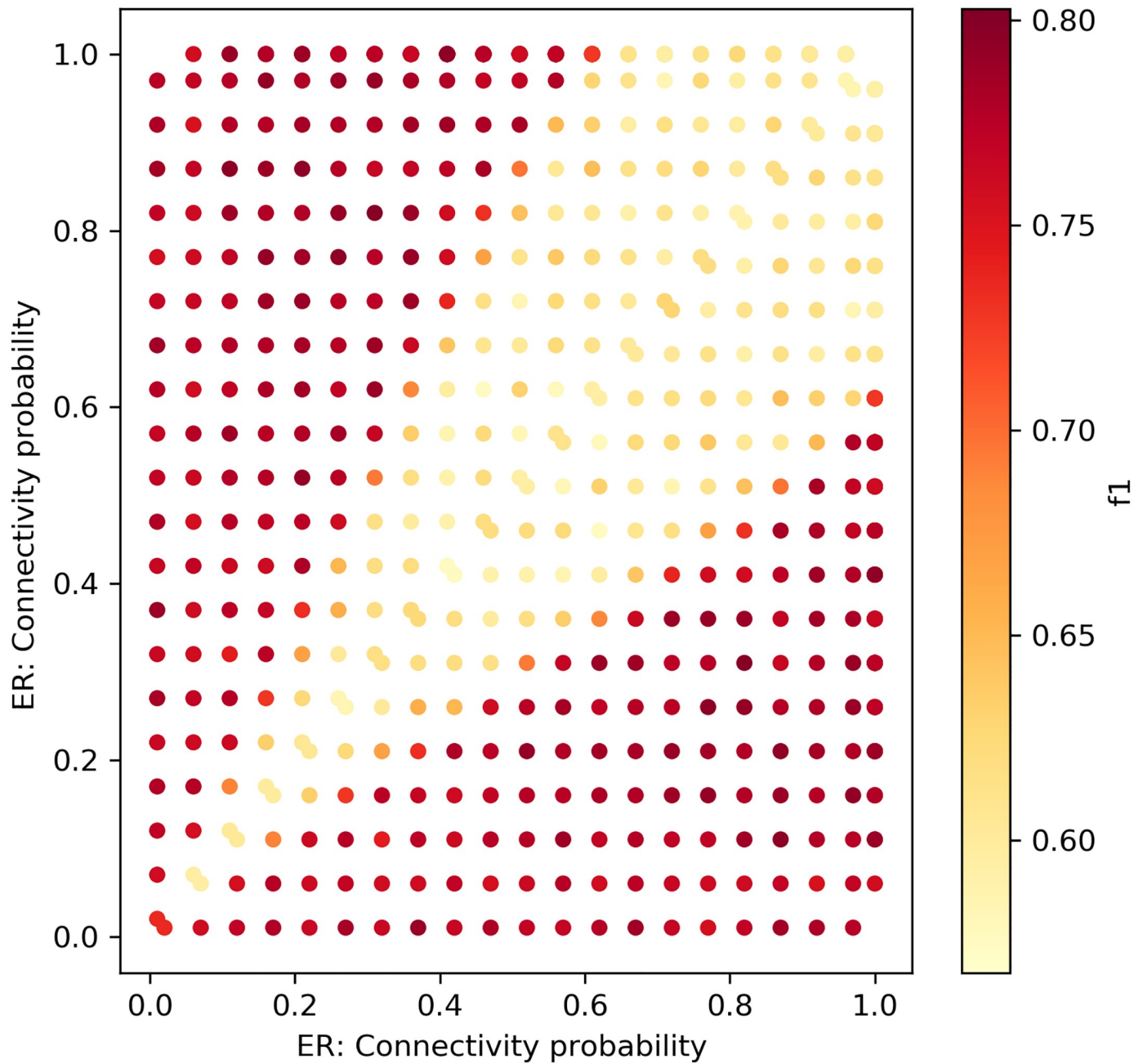

**Fig 14. CPD for random networks with RH distance.** Sensitivity test for ER networks with sizes $\sim N(\mu = 100, \sigma^2 = 10)$.

https://doi.org/10.1371/journal.pone.0231035.g014

paved the path to a further understanding of the dynamics of networks [45], and findings corroborate that structural properties have a prominent effect on the longitudinal dynamics of networks and their actors [43, 44, 46–48].

Considering that distributions in complex systems have practical importance as an aiding tool for data interpretation and event prediction [9, 10], this work explores the interplay between points of change and this fundamental structural distribution in social organizations

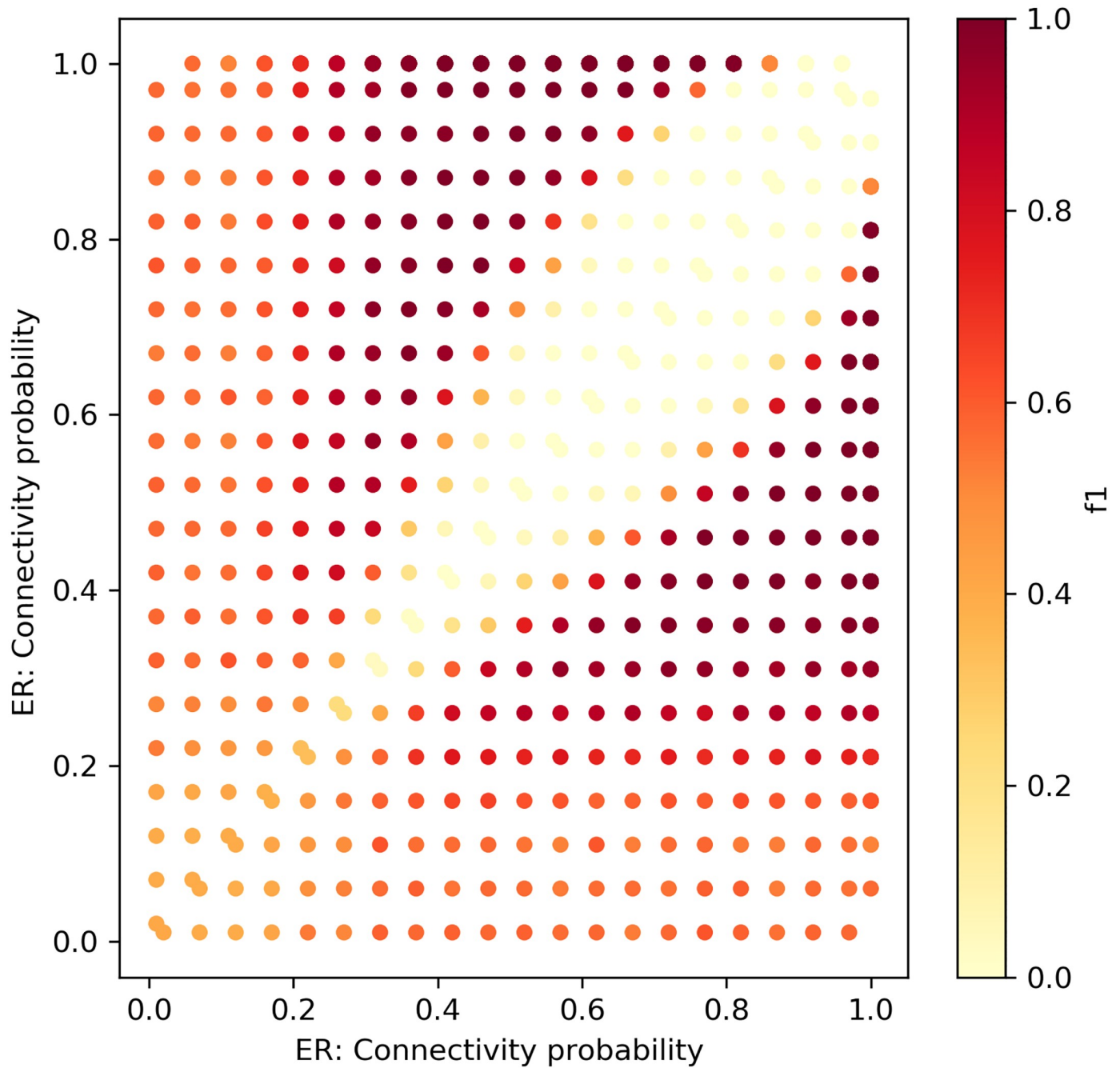

**Fig 15. CPD for random networks with KL distance.** Sensitivity test for ER networks with sizes $\sim N(\mu = 100, \sigma^2 = 10)$.

https://doi.org/10.1371/journal.pone.0231035.g015

and systems. Social interactions are dynamic, and in organizations and social venues (as is the case with the AskUbuntu forum), different subsets of the networks interact at different times. A distribution-based framework like the one presented in this work enables a variable number of nodes at each window of time and hence is size agnostic. It also does not require historical information and can be used for online detection. This is in contrast to model-based frameworks, which are limited by definition to a stringent subset of traceable interacting players

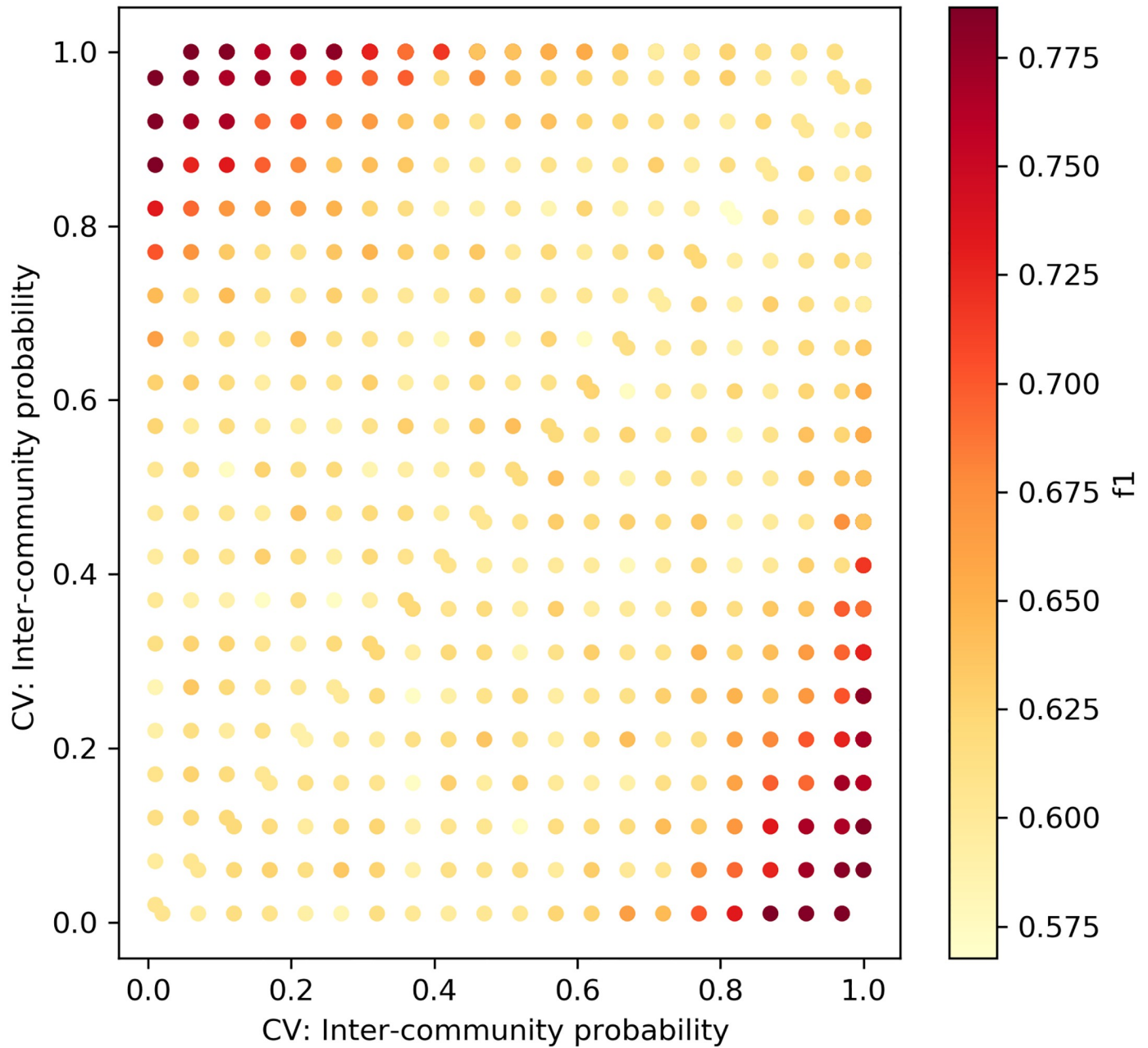

**Fig 16. CPD for caveman networks with RH distance.** Sensitivity test for Caveman networks with sizes $\sim N(\mu = 100, \sigma^2 = 10)$.

https://doi.org/10.1371/journal.pone.0231035.g016

over time. The framework presented here captures the distribution of interactions in each window of time, regardless of the participants' history. It also does not assume past correlations between the parties involved.

The framework can be employed with different distance metrics. Our results demonstrate that for moderate-size networks, the KS distance metric yields good performance, better than KL and RH. It is also widely known to be fast [36]. For networks with long-tail distributions,

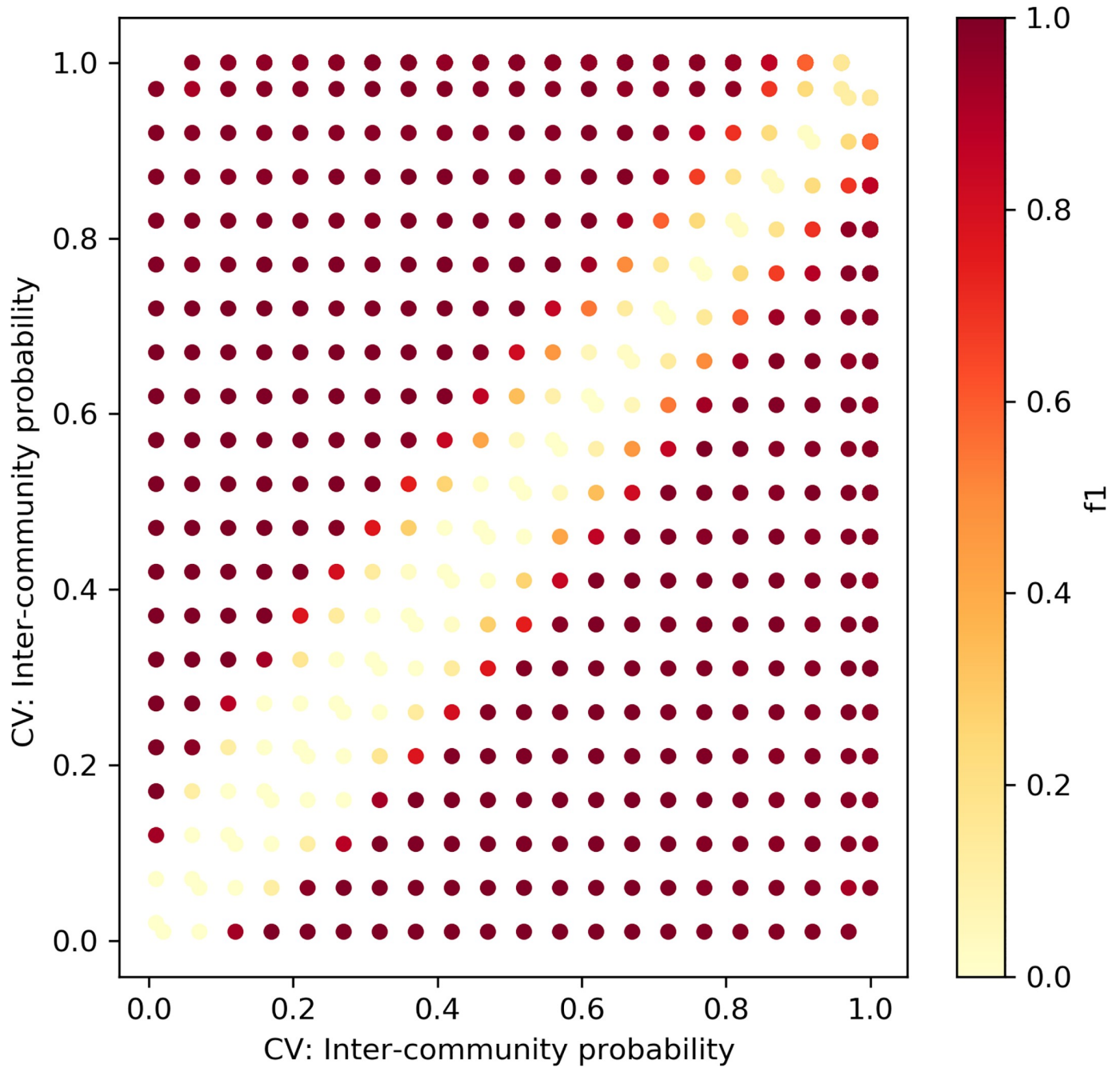

**Fig 17. CPD for caveman networks with KL distance.** Sensitivity test for Caveman networks with sizes $\sim N(\mu = 100, \sigma^2 = 10)$.

https://doi.org/10.1371/journal.pone.0231035.g017

the RH distance metric gives good results, as it is sensitive to accumulated changes in the tail [27].

## Conclusion

Our framework for size-agnostic detection of changes works across different generative models and real datasets, achieving very high recall and good precision of detection. During the

**Table 2. CPD KS framework performance summary.**

| | | Dataset | CPD framework | | GHRG [3] | |
|---|---|---|---|---|---|---|
| | | | Prec. | Recall | Prec. | Recall |
| | | Enron | 0.9 | 0.9 | 1.0 | 0.36 |

| Solution | ER sensitivity | | | | Caveman sensitivity | | | |
|---|---|---|---|---|---|---|---|---|
| | Precision | | Recall | | Precision | | Recall | |
| | Mean | Std | Mean | Std | Mean | Std | Mean | Std |
| Framework | 0.94 | 0.08 | 0.96 | 0.08 | 0.78 | 0.41 | 0.59 | 0.45 |
| RH CPD | 0.19 | 0.05 | 0.75 | 0.28 | 0.23 | 0.02 | 0.91 | 0.13 |
| KL CPD | 0.65 | 0.39 | 0.73 | 0.41 | 0.86 | 0.26 | 0.79 | 0.36 |

https://doi.org/10.1371/journal.pone.0231035.t002

work, we have identified an interesting trade-off between precision and recall of detection when considering the size of the network and detectability. We intend to study this tradeoff in future research further. An additional interesting line of research is to quantify the nature of the change in the distribution in response to different events.

## Author Contributions

**Conceptualization:** Hadar Miller, Osnat Mokryn.

**Data curation:** Hadar Miller.

**Formal analysis:** Hadar Miller, Osnat Mokryn.

**Methodology:** Hadar Miller, Osnat Mokryn.

**Software:** Hadar Miller.

**Supervision:** Osnat Mokryn.

**Writing – original draft:** Osnat Mokryn.

**Writing – review & editing:** Osnat Mokryn.